
\topmargin=.8in \vsize=9.4in
\baselinestretch=1200



\font\twelveBbb=msym10 scaled \magstep1
\font\nineBbb=msym9
\font\sevenBbb=msym7
\newfam\Bbbfam
\textfont\Bbbfam=\twelveBbb
\scriptfont\Bbbfam=\nineBbb
\scriptscriptfont\Bbbfam=\sevenBbb
\def\Bbb{\fam\Bbbfam\twelveBbb}
\def\Z{{\Bbb Z}}

\footnotenumstyle{symbols}
\footnotenum= 0

\def\n{\noindent}

\def\c#1:#2{c^{#1}_{#2}}

\def\subon{\subequationnumstyle{alphabetic}}
\def\suboff{\subequationnumstyle{blank}}
\def\h{\hbox to .5cm{\hfill}}
\def\hof{\hbox to .15cm{\hfill}}
\def\htf{\hbox to .35cm{\hfill}}

\def\t{\theta}
\def\bt{\bar\theta}

\def\undbib{{\underline{\hbox to 2.5cm{\hfill}}},\ }

{
\pagenumstyle{blank}
\hbox to 1cm{\hfill}
\vskip .3 truecm
{\rightline{\hbox to 4.5cm{\vtop{\hsize= 4.5cm
\baselinestretch=960\footnotefonts
\hfill\\
CALT-68-1757\\
DOE RESEARCH AND\\
DEVELOPMENT REPORT}}{\hbox to .35cm{\hfill}}}}
\vskip 2.0 truecm

\centerline{{\bf\bigfonts Comments on Fractional Superstrings}\footnote{Work
supported in part
by the U.S. Department of Energy under Contract no. DEAC-03-81ER40050.}}
\vskip 1.4 truecm
\centerline{Gerald B. Cleaver}
\vskip .7 truecm
\centerline{{\it California Institute of Technology, Pasadena,} CA, 91125}
\vskip 1.5 truecm
\centerline{\bf Abstract}
\vskip .3 truecm
\parindent=.5cm\narrower
Using the factorization
approach of Gepner and Qiu, I systematically rederive
the closed fractional superstring partition functions for
$K=4,\, 8,\, {\rm ~and~} 16$.
For these theories the relationship between
the massless graviton and gravitino sector and the purely massive
sectors is explored.
Properties of the massive sectors are investigated.
A twist current in these models
is found responsible for the occurrence of $N=1$ space-time supersymmetry.
I show this twist current transforms bosonic (fermionic)
projection states into
fermionic (bosonic) non-projection states and vice-versa.\smallskip
\vskip .5truein
\noindent
\it To appear in the Proceedings of the International Workshop on
String Theory, Quan-
\hbox to .05cm{\hfill}tum Gravity and the Unification of
Fundamental Interactions, 21-26 September 1992,

\centerline {Rome, Italy.}
\hfill\vfill\eject}

\pagenumstyle{arabic}\pagenum=0
\footnotenum=0
\sectionnumstyle{blank}
\section{\bf Section 1: Brief Review of Fractional Superstrings}
\sectionnumstyle{arabic}\sectionnum=1

In the prior talk, Keith Dienes presented a general review of fractional
superstrings with critical dimensions of six, four, and three.
These fractional superstrings are based on $SU(2)_K/U(1)$ coset (a.k.a.
parafermion) conformal field theories.
The relationship  between critical dimension and the level of the
$SU(2)_K/U(1)$ theory is,
$$D= 2+ 16/K\,\, . \eqno\eqnlabel{dim1}\,\, .$$
Thus
integer\footnote{In ref. \putref{aspects}
we have shown that
non-integer $D$ fractional superstrings cannot have space-time
supersymmetry.} $D= 6$, $4$, $3$ corresponds to $K= 4$, $8$, $16$,
respectively. In the
present talk, I will focus specifically on the associated partition
functions, how we can derive these from ``first principles'', and what is
learned from this.  These closed string partition functions
$Z({\rm level~} K)$
have the
general form $$ Z({\rm level~} K) = \vert A_K\vert^2 + \vert B_K\vert^2 + \vert
C_K\vert^2\,\, .
\eqno\eqnlabel{pf1}$$
The $A_K$ term in eqn. (\puteqn{pf1}) contains the massless graviton and
gravitino.  These $D<10$ fractional superstrings have a new feature not
present in the standard $D=10$
superstrings (which correspond to K=2). This is the existence of the
massive $B_K$ and $C_K$ sectors.  These additional
sectors were originally derived by the authors of refs.
\putref{ref63,ref62,ref65} by applying $S$ transformations
to the $A_K$ sector and then demanding modular invariance of the
theory.  I will remark (1) on new aspects of the relationship between
the $B_K$ and $C_K$ sectors and the $A_K$ sector
and (2) on the
presence of $N=1$ spacetime supersymmetry (susy) in all sectors.  I will
demonstrate that $N=1$ susy results from
the action of a twist current used in the
derivation of these partition functions.  Only by this twisting can
cancellation between bosonic and fermionic terms occur at each mass level
in the $A_K$ and $B_K$ sectors.  The same twisting
results a in
``self-cancellation'' of terms in the $C_K$ sector (which exists only in
the four and three dimensional models).  This self-cancellation
may suggest an anyonic
interpretation of the $C_K$ sector states.\footnote{Uncompactified space-time
anyons can presumably exist only in three or less dimensions.
This would seem to
contradict a claim that the $D=4$ model may contain space-time anyons. I
will argue shortly that one dimension of the $D=4$ fractional string is
probably compactified. Examination of the $B_K$ sector in the $D=4$ model
further suggests this. Anyonic interpretation of the $C_K$ sector fields was
first proposed in ref.
\putref{ref63}.}

Before I systematically derive the fractional superstring partition functions
for each critical dimension, I
will review the partition function $Z(\phi^j_m(z))$ for the Verma
module, $[\phi^j_m]$,\footnote{From here on I do not distinguish between
the primary field $\phi^j_m$ and its complete Verma module $[\phi^j_m]$.
Instead, $\phi^j_m$ represents both where meaningful.}
containing a single (holomorphic) parafermionic primary field
$\phi^j_m(z)$ and its descendents.  $$ Z(\phi^j_m(z))=
\eta(\tau)c^{2j}_{2m}(\tau)\, ,\eqno\eqnlabel{2a}$$ where $\eta$ is the
Dedekind eta-function, $$ \eta= q^{1/24}\prod^{\infty}_{n=1}(1-q^n)
\eqno\eqnlabel{defeta}$$ with $q= e^{2\pi i \tau}$, and $c^{2j}_{2m}$ is a
string function\markup{\putref{ref8}} defined by
$$ c^{2j}_{2m} = {1\over {\eta^{-3}(\tau)}} \sum_{x,y}sign(x) q^{x^2(K+2) -
y^2K} \eqno\eqnlabel{cfna}$$
where in (\puteqn{cfna}) the conditions\\
\n {1.} $ -\vert x\vert <y\leq\vert x\vert $,\\
\n {2.} either $x= {l+1\over 2(K+2)}\pmod{1}$
or $({1\over 2} -x) = {l+1\over 2(K+2)}\pmod{1}$; and\\
\n {3.} either $y= {n\over 2K} \pmod{1}$ or $({1\over 2} + y)= {n\over
2K}\pmod{1}$

\noindent must be met simultaneously.\markup{\putref{ref63}} These
string functions obey the same equivalences as their associated primary
fields $\phi^j_m$:
\subequationnumstyle{alphabetic}
$$c^{2j}_{2m}= c^{2j}_{2m+2K} = c^{K-2j}_{2m-K}\, .\eqno\eqnlabel{cid-a}$$
Additionally $$c^{2j}_{2m}= c^{2j}_{-2m}\, .\eqno\eqnlabel{cid-b}$$
\subequationnumstyle{blank}

For each space-time dimension in these theories, a partition function
term of the form (\puteqn{2a})
is tensored with the partition function $Z\left(X(z)\right)$
for an uncompactified chiral boson $X(z)$.
$$ Z\left(X(z)\right)\propto {1\over\eta(\tau)}.
\eqno\eqnlabel{partstdim}$$
Thus the $\eta(\tau)$ factors cancel out in
$Z(\phi^j_m(z))\times Z(X(z))$. Similar cancellation of
$\bar\eta(\bar\tau)$ occurs in the antiholomorphic sector.
In the following partition functions I suppress the additional
trivial net factor of
$(Im\, \tau)^{-8/K}$
contributed by the $D-2$ holomorphic and anti-holomorphic
world sheet boson partition functions together.
\sectionnumstyle{blank}
\section{\bf Section 2: Fractional Superstring Partition Functions}
\sectionnumstyle{arabic}\sectionnum=2\equationnum=0

By the string function equivalences, the partition functions for the
level $K$ fractional superstrings in refs.
\putref{ref63,ref62,ref65,ref12}
in critical space-time dimensions
\hbox {$D= 2 + {16\over K}= 10, 6,$}
{$\,4,\, {\rm~and~}3$}
can be written (in light-cone gauge) as:

{\settabs 8 \columns
\+ \cr
\+ $D=10$ & $(K=2)$: &&$Z =  \vert A_2\vert^2$, ~where\cr}
$$\eqalignno{ A_2 &= {1\over 2}\left\{ (c^0_0 + c^2_0)^8 - (c^0_0 -
c^2_0)^8 \right\}_{boson} - 8(c^1_1)^8_{fermion}\cr &= 8\left\{ (c^0_0)^7
c^2_0 + 7(c^0_0)^5(c^2_0)^3 +7(c^0_0)^3(c^2_0)^5 +
c^0_0(c^2_0)^7\right\}_{boson}  -
8(c^1_1)^8_{fermion} &\eqnlabel{part2}}$$ {\settabs 8 \columns
\subon
\+ $D=6$ & $(K=4)$: &&$Z= \vert A_4\vert^2 + 3\vert B_4\vert^2$, ~where\cr}
$$\eqalignno {A_4 &= {\rm\hskip .37 truecm}4\left\{(c^0_0 + c^4_0)^3
(c^2_0) - (c^2_0)^4\right\}\cr &\quad + 4\left\{(c^0_2 + c^4_2)^3 (c^2_2) -
(c^2_2)^4\right\}&\eqnlabel{part4-a}\cr B_4 &= {\rm\hskip
.37truecm}4\left\{(c^0_0 +
c^4_0)(c^0_2+c^4_2)^2(c^2_0) - (c^2_0)^2(c^2_2)^2\right\}\cr &\quad +
4\left\{(c^0_2 + c^4_2)(c^0_0+c^4_0)^2(c^2_2) -
(c^2_2)^2(c^2_0)^2\right\}&\eqnlabel{part4-b}\cr}$$
{\settabs 8 \columns
\suboff
\subon
\+ $D=4$ & $(K=8)$:
&&$Z= \vert A_8\vert^2 + \vert B_8\vert^2 + 2\vert C_8\vert^2$, ~where\cr}
$$\eqalignno { A_8 & = {\rm\hskip .37 truecm}2\left\{(c^0_0 + c^8_0)(c^2_0
+ c^6_0) - (c^4_0)^2\right\}\cr &\quad +2\left\{(c^0_4+ c^8_4)(c^2_4+c^6_4)
- (c^4_4)^2\right\}&\eqnlabel{part8-a}\cr B_8 &= {\rm\hskip .37
truecm}2\left\{(c^0_0+c^8_0)(c^2_4+c^6_4)-(c^4_0c^4_4)\right\}\cr &\quad +
2\left\{(c^0_4+c^8_4)(c^2_0+c^6_0)-(c^4_4c^4_0)\right\}&\eqnlabel{part8-b}\cr
C_8 &= {\rm
\hskip .37truecm}2\left\{(c^0_2 + c^8_2)(c^2_2 + c^6_2) -
(c^4_2)^2\right\}\cr &\quad +2\left\{(c^0_2 + c^8_2)(c^2_2 + c^6_2) -
(c^4_2)^2\right\}&\eqnlabel{part8-c}\cr}$$ {\settabs 8 \columns
\suboff
\subon
\+ $D=3$ & $(K=16)$:
&&$ Z = \vert A_{16}\vert^2 + \vert C_{16}\vert^2 $, ~where\cr}
$$\eqalignno{A_{16} &= {\rm\hskip .3 truecm}\left\{(c^2_0 + c^{14}_0) -
c^8_0\right\}\cr &\quad +\left\{(c^2_8 + c^{14}_8) - c^8_8\right\}
&\eqnlabel{part16-a}\cr
C_{16} &= {\rm\hskip .3 truecm}\left\{(c^2_4 + c^{14}_4) -
c^8_4\right\}\cr &\quad +\left\{(c^2_4 + c^{14}_4) -
c^8_4\right\}\,\, .&\eqnlabel{part16-b}\cr}$$
\suboff

The $D=10$ partition function, written in terms of string functions,
was originally
obtained by the authors of refs. \putref{ref63,ref65} using the
equivalences between the $K=2$ string functions and the non-zero Jacobi
$\theta$-functions.\footnote{The $K=2$ parafermion $c={1\over 2}$ CFT
corresponds to an Ising (free fermion) model.}
The partition functions for $K>2$ were derived demanding
a massless spin two particle, modular invariance, and no tachyons.
In each model, the massless spin-2 particle and its supersymmetric partner
arise from the $A_K$ sector. The $B_K$ and $C_K$ sectors were obtained by
acting on the $A_K$ sector with
the $SL(2,\Z)$ modular group generators, $S: \tau\rightarrow -1/\tau$,
and $T:\tau\rightarrow \tau +1$.
At each level $K$, every mod-squared term is individually zero.  This is
consistent with $N=1$ susy in flat space-time and suggests cancellation
between bosonic and fermionic terms at each mass level.
This leads to the following
identities\markup{\putref{ref63}}:
$$A_2=A_4=B_4=A_8=B_8=C_8=A_{16}=C_{16}=0\,\, .\eqno\eqnlabel{partident}$$

The factorization method of Gepner and Qiu\markup{\putref{ref7}} for
string function partition functions allow us to rederive the above
partition functions systematically.  From this approach we can express a
general parafermionic partition function (with the level $K$
of the string functions often henceforth suppressed),
\subequationnumstyle{alphabetic}
$$ Z= \vert \eta\vert^2 \sum N_{l,n,\bar l,\bar n} c^l_n \bar c^{\bar
l}_{\bar n}\, ,\eqno\eqnlabel{partfn2-a}$$ as $$Z= \vert\eta\vert^2\sum
{1\over 2}L_{l,\bar l}M_{n,\bar n}c^l_n \bar c^{\bar l}_{\bar n}\,
,\eqno\eqnlabel{partfn2-b}$$
\subequationnumstyle{blank}
(with $c^{l=2j}_{n=2m}=0$ unless $l-n\in 2\Z$ since $\Phi^j_m=0$ for
$j-m\not\in\Z$).  As a result of the
$$N_{l,n,\bar l,\bar n} = {1\over 2}L_{l,\bar l}\, M_{m,\bar m}
\eqno\eqnlabel{Nfactorization}$$
factorization,
we can construct all modular invariant
partition functions
(MIPF's) for
para-
fermions
from a tensor product
of modular invariant solutions for the $(l,\bar l)$ and
$(n,\bar n)$ indices separately. This results from the
definition of a level $K$ string functions $c^l_n$ in terms of the
$SU(2)_K$ characters $\chi_l$ and the Jacobi theta function
$\theta_{n,K}$:\footnote{The associated relationship between the level-$K$
$SU(2)$ primary fields $\Phi^j$ and the parafermionic $\phi^j_m$ is
$$\Phi^j= \sum_{m=-j}^j \phi^j_m\,  :exp\left\{ {i{m\over \sqrt{K}}
\varphi}\right\}:$$
where $\varphi$ is the $U(1)$ boson field of the $SU(2)$ theory.}
$$\chi_l(\tau)
= \sum^K_{n= -K+1} c^l_n(\tau)\theta_{n,K}(\tau)\,
,\eqno\eqnlabel{partfn4}$$ where the theta function is defined by
$$\theta_{n,K}(\tau) = \sum_{p\in\Z + {n\over 2K}}e^{2\pi i K p^2\tau}\,
,\eqno\eqnlabel{thetafn}$$ and $\chi_l$ is the character for the spin ${1\over
2}l$ representation of $SU(2)_K$, $$\chi_l(\tau) =
{\theta_{l+1,K+2}(\tau)-\theta_{-l-1,K+2}(\tau)\over
\theta_{1,2}(\tau)-\theta_{-1,2}(\tau)}
\, .\eqno\eqnlabel{chifn}$$
This factorization is clearly seen in the transformation properties of
$c^l_n$ under the modular group generators $S$ and
$T$,
\subequationnumstyle{alphabetic}
$$\eqalignno{S: c^l_n &\rightarrow {1\over \sqrt{-i\tau K(K+2)}}
\sum_{l'=0}^{K}\sum_{n'= -K+1 \atop l'-n`\in 2Z}^K exp\left\{{i\pi n
n'\over K}\right\} sin\left\{{\pi (l+1)(l'+1)\over K+2}\right\} c^{l'}_{n'}
{\rm ~~~~~~~~~} &\eqnlabel{ctrans-a}\cr T: c^l_n &\rightarrow exp\{2\pi
i[{j(j+1)\over K+2} - {m^2\over K} - {K\over 8(K+2)}]\} c^l_n\,\, .
&\eqnlabel{ctrans-b}\cr}$$
\subequationnumstyle{blank}
Thus, eqn. (\puteqn{partfn2-b}) is modular invariant if and only if the
$SU(2)$ affine partition function
$$ W=
\sum_{l,\bar l= 0}^K L_{l,\bar l}\chi_l(\tau)\bar\chi_{\bar
l}(\bar\tau)\eqno\eqnlabel{partfn3}$$
and the $U(1)$ partition function
$$ V= {1\over\vert\eta(\tau)\vert^2}
\sum_{n,\bar n= -K+1}^K M_{n,\bar n}\theta_{n,K}\bar\theta_{\bar
n,K}\eqno\eqnlabel{partfn10}$$ are simultaneously modular invariant.
That is $N_{l,n,\bar l,\bar n}= {1\over 2}L_{l,\bar l}M_{n,\bar n}$ corresponds
to a
MIPF (\puteqn{partfn2-a}) if and only if $L_{l,\bar l}$ and $M_{n,\bar n}$
correspond to MIPF's of the forms (\puteqn{partfn3}) and
(\puteqn{partfn10}), respectively.

This factorization is also possible for parafermion tensor product theories,
with matrices $\bf L$ and $\bf M$ generalized to tensors.  Any
tensor $\bf M$ corresponding to a MIPF for $p$ factors of $U(1)$
CFT's can be written as a tensor product of $p$ independent matrix $\bf M$
solutions to (\puteqn{partfn10}) twisted by simple
currents $\cal J$.\markup{\putref{me}}
This approach greatly simplifies the derivation of
the fractional superstring partition functions,
while simultaneously suggesting much about the meaning
of the different sectors, the origin of space-time supersymmetry and
related ``projection'' terms.

\subsectionnumstyle{blank}
\subsection{{\bf 2.1}: \it Affine Factor and ``W'' Partition Function}
\subsectionnumstyle{blank}

In the $A_K$ sectors defined by eqns.
(\puteqn{part4-a}, \puteqn{part8-a}, \puteqn{part16-a})
the terms inside the first (upper) set of brackets,
which carry ``$n\equiv 2m=0$'' subscripts, correspond to space-time bosons
and the terms inside the second (lower) set, carrying ``$n= K/2$''
correspond to space-time fermions.  Expressing the $A_K$ sectors in this form
makes a one--to--one correspondence between bosonic and fermionic states in the
$A_K$ sector manifest. If I remove the subscripts on the string functions
in the bosonic and fermionic subsectors
(which is parallel to replacing $c^l_n$ with
$\chi_l$) we find the subsectors become equivalent. In fact, under this
operation of removing the ``$n$'' subscripts
(which I will denote by $\buildrel {\rm
affine}\over\Longrightarrow$), all sectors become equivalent up to an
integer coefficient:\hfill\vfill\newpage
\subon
{\settabs 8 \columns\+ $D=6$ &$(K=4)$:\cr} $$A_4,B_4 \hbox to
1cm{\hfill}{\buildrel {\rm
affine}\over\Longrightarrow}\hbox to 1cm{\hfill}
A_4^{affine}\equiv (\chi_0+\chi_K)^3\chi_{K/2}-(\chi_{K/2})^4
\eqno\eqnlabel{affine-a}$$
{\settabs 8 \columns\+ $D=4$ &$(K=8)$:\cr} $$A_8,B_8,C_8
\hbox to 1cm{\hfill}{\buildrel {\rm affine}\over\Longrightarrow}
\hbox to 1cm{\hfill}
A_8^{affine}\equiv (\chi_0+\chi_K)(\chi_2+\chi_{K-2})
- (\chi_{K/2})^2 \eqno\eqnlabel{affine-b}$$ {\settabs 8 \columns\+ $D=3$
&$(K=16)$:\cr} $$A_{16},C_{16}
\hbox to 1cm{\hfill}{\buildrel {\rm affine}\over\Longrightarrow}\hbox to
1cm{\hfill}
A_{16}^{affine}\equiv (\chi_2 + \chi_{K-2}) -\chi_{K/2}\,\,
.\eqno\eqnlabel{affine-c}$$
\suboff
Eqns. (2.13a-c)) all have the
general form $$ A_K^{affine} \equiv (\chi_0 +\chi_K)^{D-3}(\chi_2+\chi_{K-2}) -
(\chi_{K/2})^{D-2}\,\, . \eqno\eqnlabel{affall}$$
Thus,
$$ W({\rm level~} K)=
\vert A^K_{aff}\vert^2\, .\eqno\eqnlabel{affine2}$$ (Note that the modular
invariance of $W$ requires that $A_K^{aff}$ transforms back into itself under
$S$.)

The class of partition functions (\puteqn{affine2}) is indeed modular
invariant and possesses special qualities. This is easiest to show for
$K=16$.  The $SU(2)_{16}$ MIPF's for $D=3$ are trivial to classify since
at this level the A--D--E classification forms a complete basis set of modular
invariants, even for MIPF's containing terms with negative coefficients. The
only free parameters in $K=16$ affine partition functions $Z\left(
SU(2)_{16}\right)$ are integers $a$, $b$, and $c$ where $$Z\left(
SU(2)_{K=16}\right) =
a\times Z({\rm A}_{17})+b\times Z({\rm D}_{10})+c\times Z({\rm E}_7)\,\,
.\eqno\eqnlabel{affine3}$$

Demanding that a
tachyonic state not be in the Hilbert space of states in the $K=16$
fractional superstring with intercept $v= c/24$, defined by $$L_0\vert
physical\rangle = v\vert physical\rangle\, ,
\eqno\eqnlabel{intercept}$$
removes these degrees of freedom and requires $a= -(b+c)=0$,
independent of the possible $(n,\bar n)$ partition
functions.
These specific values for $a$, $b$, and $c$ give us (\puteqn{affine2})
for this level:
$$W\left( K=16 \right) = Z({\rm D}_{10})-Z({\rm E}_{7})
= \vert A^{aff}_{16}\vert^2
\,\, .\eqno\eqnlabel{affine4}$$

Though not quite as straightforward a process,
we can also derive the affine partition functions $W(K)$ for the
remaining levels.  The affine factors in the
\hbox {$K=4 {\rm ~and~} 8$}
partition
functions involve twisting by a non-simple current.\footnote{Models
involving
non-simple current twisting is discussed briefly in the appendix.
See also refs. \putref{ref10} and \putref{thesis}
for a discussion of simple and non-simple
currents and their ``twisting'' effects.
Multiloop modular invariance of $SU(2)_K$ tensor product models with
non-simple current twisting is an issue currently under my investigation.}
These cases
correspond to theories that are the difference between a
${\bigotimes\atop {D-2\atop {\rm factors}}} {\rm D}_{{K\over 2} +2}$ tensor
product model
and a ${\bigotimes\atop{D-2\atop {\rm factors}}} {\rm D}_{{K\over 2} +2}$
tensor
product model twisted by the affine
current\footnote{I
have left off the space-time indices on most of the following currents
and fields.  I am working in light-cone gauge so only indices for
transverse modes are implied.
The $D-2$ transverse dimensions are assigned
indices in the range 1 to $D-2$ (and are generically represented by
lowercase Greek superscripts.) When space-time indices
are suppressed,
the fields and their corresponding partition
functions acting along directions 1 to $D-2$
are ordered from left to right, respectively, for both
the holomorphic and antiholomorphic sectors separately.
Often I will be still more implicit
in my notation and will express $r$ identical factors of $\phi^j_m$ along
consecutive directions (when these directions are either all compactified or
uncompactified) as $(\phi^j_m)^r$. Thus, eqn. (2.21) for $K=8$ should be
read as
$$J_{non-simple}^{K=8}\equiv
(\phi^{K/4}_0)^{\mu=1}(\phi^{K/4}_0)^{\nu=2}(\bar\phi^1_0)^{\bar\mu=1}
(\bar\phi^0_0)^{\bar\nu=2}\,\, .$$}
$$J_{non-simple}^{K,affine}=(\Phi^{K\over
4})^{D-2}\bar\Phi^1(\bar\Phi^0)^{D-3}\,\, .\eqno\eqnlabel{jkaff}$$
The equivalent parafermionic twist current is obvious,
$$J_{non-simple}^{K,parafermion}=(\phi^{K\over 4}_0)^{D-2}\bar\phi^1_0
(\bar\phi^0_0)^{D-3}\,\, .\eqno\eqnlabel{affine11}$$ (This derivation
applies to
the $K=16$ case also .)

\subsection{{\bf 2.2}: \it Theta Function Factor and the ``$V$'' Partition
Function}

I now consider the theta function factors
carrying the indices $(n,\bar n)$ in the fractional superstring partition
functions.
Since all
$A_K$, $B_K$, $C_K$ sectors in the level $K$ fractional superstring partition
function (and even the boson and fermion subsectors separately in $A_K$)
contain the
same affine factor, it is clearly the choice (or lack thereof after
elimination of tachyons) of the theta function
factor which determines the level of space-time supersymmetry of the
fractional superstring theories. That is, space-time spins of particles in the
Hilbert space of states depend upon the
${\bf M}'s$ that are allowed in tensored versions of eqn.
(\puteqn{partfn10}).  In the case of matrix $\bf M$ rather than a more
complicated tensor, invariance of (\puteqn{partfn10})
under $S$ requires that the components $M_{n\bar n}$
be related by
\subon
$$ M_{n',\bar n'} = {1\over 2K}\sum_{n,\bar n}M_{n,\bar n} e^{i\pi n
n'/K}e^{i\pi \bar n \bar n'/K}\,\, , \eqno\eqnlabel{m-a}$$
and $T$ invariance
demands that $$ {n^2 - \bar n^2\over 4K}\in\Z\,\, ,
{\rm ~~if~} M_{n,\bar n}\neq 0\,\,. \eqno\eqnlabel{m-b}$$
\suboff
At every level $K$ there is a unique modular
invariant function corresponding to
each factorization\markup{\putref{ref7}},
$\alpha\times\beta=K$, where $\alpha,\,\, \beta\in\Z$.
Denoting the matrix elements of ${\bf M}^{\alpha,\beta}$ by
$M^{\alpha,\beta}_{n,\bar n}$,
they are given by\footnote{By eqn. (2.23), $M^{\alpha,\beta}_{n,\bar n}
= M^{\beta,\alpha}_{n,-\bar n}$.
Hence, ${\bf M}^{\alpha,\beta}$ and ${\bf M}^{\beta,\alpha}$
result in equivalent fractional superstring partition functions.
To avoid this redundancy, I choose $\alpha\leq\beta$.

Throughout this subsection
I will view the $n$ as representing, simultaneously,
the holomorphic $\theta_{n,K}$ characters for $U(1)$ theories
and, in some sense,
the holomorphic string functions, $c^0_{n}$, for parafermions.
($\bar n$ represents the antiholomorphic parallels.)
However, I do not intend to imply that the string functions
can actually be factored into $c^l_0\times c^0_n=c^l_n$.  Rather,
I mean to use this in eqns. (2.25b, 2.27b, 2.29b) only as an
artificial construct for
developing a deeper understanding of the
function of the parafermion primary fields (Verma modules) $\phi^0_m$ in
these models.  In the case of the primary fields, $\phi^j_m$,
factorization is, indeed, valid for integer $j$ and $m$:
$\phi^j_0\otimes\phi^0_m=\phi^j_m\,\, .$}
$$M^{\alpha,\beta}_{n,\bar n} = {1\over 2}
\sum_{x\in\Z_{2\beta}\atop y\in\Z_{2\alpha}}
\delta_{n,\alpha x +\beta y}\delta_{\bar n,\alpha x -\beta y}\,\, .
\eqno\eqnlabel{m-c}$$

Thus,
for $K=4$ the two distinct choices for the matrix ${\bf M}^{\alpha,\beta}$
are ${\bf M}^{1,4}$ and ${\bf M}^{2,2}$; for
$K=8$, we have ${\bf M}^{1,8}$ and ${\bf M}^{2,4}$; and
for $K=16$, the three alternatives are ${\bf M}^{1,16}$,
${\bf M}^{2,8}$, and ${\bf M}^{4,4}$.
${\bf M}^{1,K}$ represents the level $K$ diagonal, $n=\bar n$,
partition function.
${\bf M}^{\alpha, \beta={K\over\alpha}}$
corresponds to the
diagonal partition function twisted by a $\Z_{\alpha}$ symmetry.
(Twisting by $\Z_{\alpha}$ and $\Z_{K/\alpha}$ produce isomorphic
models.)
Simple
tensor products of these ${\bf M}^{\alpha,\beta}$ matrices are insufficient
for
producing fractional superstrings with $N=1$ space-time susy (and, thus, no
tachyons).
I have found that twisting by a special simple current is required to
achieve this.
Of the potential choices for the $U(1)$ MIPF's,
$V({\rm level~} K)$, the
following are the only ones that produce numerically zero
fractional superstring partition functions:
{\settabs 8 \columns
\+ $D=6$ & $(K=4)$:\cr
\+ \cr}

The ${\bf M}= {\bf M}^{2,2}\otimes{\bf M}^{2,2}\otimes{\bf
M}^{2,2}\otimes{\bf M}^{2,2}$ model twisted by the
simple current\footnote[$\# $]{The parafermion primary fields $\phi^0_m$
have simple fusion rules,
$$\phi^0_m\otimes\phi^0_{m'}=\phi^0_{m+m'} \pmod{K}$$
and form a $Z_K$ closed subalgebra.  This fusion rule, likewise, holds for
the $U(1)$ fields $:exp\{i{m\over K}\varphi\} :.$  This isomorphism
makes it clear that any simple current, ${\cal J}_K$, in this subsection can be
expressed equivalently either in terms of these parafermion fields
or in terms of
$U(1)$ fields.  In view of the following discussion, I define all of the
simple twist currents, ${\cal J}_K$, as composed of the former.}

$${\cal J}_4\equiv \phi_{K/4}^0\phi_{K/4}^0\phi_{K/4}^0\phi_{K/4}^0
\bar\phi_0^0\bar\phi_0^0\bar\phi_0^0\bar\phi_0^0 \eqno\eqnlabel{j4}$$
results in the following $(n,\bar n)$ partition functions:
\subon
$$\eqalignno{V\left(K=4\right) &= \hbox to
.25cm{\hfill}[(\t_{0,4}+\t_{4,4})^4(\bt_{0,4}
 + \bt_{4,4})^4 +
(\t_{2,4} +\t_{-2,4})^4(\bt_{2,4}+ \bt_{-2,4})^4\cr
&\hbox to 1em{\hfill}
+ (\t_{0,4} +\t_{4,4})^2(\t_{2,4}+\t_{-2,4})^2(\bt_{
0,4} +\bt_{4,4})^2(\bt_{2,4} + \bt_{-2,4})^2\cr
&\hbox to 1em{\hfill}
+ (\t_{2,4}+\t_{-2,4})^2(\t_{0,4}+\t_{4,4})^2(\bt_{2,4} +\bt_{-2,4})^2
(\bt_{0,4} + \bt_{4,4})^2]_{untwisted}\cr
&&\eqnlabel{nn4-a}\cr
&\hbox to 1em{\hfill}
+ [(\t_{2,4}+\t_{-2,4})^4(\bt_{0,4}+\bt_{4,4})^4 + (\t_{4,4}+\t_{0,4})^4
(\bt_{2,4}+\bt_{-2,4})^4\cr
&\hbox to 1em{\hfill}
+ \hbox to
.15cm{\hfill}(\t_{2,4}+\t_{-2,4})^2(\t_{4,4}+\t_{0,4})^2(\bt_{0,4}+\bt_{4,4})^2
(\bt_{2,4}+\bt_{-2,4})^2\cr
&\hbox to 1em{\hfill}
+ \hbox to .15cm{\hfill}(\t_{4,4}+\t_{0,4})^2(\t_{2,4}+\t_{-2,4})^2(\bt_{2,4} +
\bt_{-2,4})^2(\bt_{0,4} +\bt_{4,4})^2]_{twisted}\,\, .}$$

Writing this in parafermionic form, and then
using string function identities
and regrouping according to $A_4$ and $B_4$ components,
results in
$$Z({\rm theta~ factor,~} K=4)= \vert (c^0_0)^4 +
(c^0_4)^4\vert^2_{_{(A_4)}} + \vert (c^0_0)^2(c^0_4)^2 + (c^0_4)^2(c^0_0)^2
\vert^2_{_{(B_4)}}\,\, .
\eqno\eqnlabel{nn4-b}$$
\suboff

{\settabs 8\columns
\+ \cr
\+ $D=4$ & $(K=8)$:\cr
\+\cr}

The ${\bf M}= {\bf M}^{2,4}\otimes{\bf M}^{2,4}$ model twisted by the
simple current $${\cal J}_8\equiv
\phi_{K/4}^0\phi_{K/4}^0\bar\phi_0^0\bar\phi_0^0\eqno\eqnlabel{j8}$$
results in
\subon
$$\eqalignno{V\left(K=8\right) &= \hbox to
.35cm{\hfill}[(\t_{0,8}+\t_{8,8})(\bt_{0,8}
+ \bt_{8,8}) +
(\t_{4,8}+\t_{-4,8})(\bt_{4,8}+\bt_{-4,8})]^2_{untwisted}\cr
&\hbox to 1em{\hfill}
+ [(\t_{2,8}+\t_{-6,8})(\bt_{2,8}+\bt_{-6,8})
+(\t_{-2,8}+\t_{6,8})(\bt_{-2,8}+\bt_{6,8})]^2_{untwisted}\cr
&&\eqnlabel{nn8-a}\cr
&\hbox to 1em{\hfill}
+ [(\t_{4,8}+\t_{-4,8})(\bt_{0,8} + \bt_{8,8})
+ (\t_{0,8}+\t_{8,8})(\bt_{4,8} + \bt_{-4,8})]^2_{twisted}\cr
&\hbox to 1em{\hfill} +
[(\t_{6,8}+\t_{-2,8})(\bt_{2,8}+\bt_{-6,8})
+(\t_{2,8}+\t_{-6,8})(\bt_{-2,8}  +\bt_{6,8})]^2_{twisted}\,\, .}$$
Hence,
$$ Z({\rm theta~factor,~} K=8)=  \vert (c^0_0)^2 +
(c^0_4)^2\vert^2_{_{(A_8)}} + \vert (c^0_0)(c^0_4)
+(c^0_4)(c^0_0)\vert^2_{_{(B_8)}} + 4\vert
(c^0_2)^2\vert^2_{_{(C_8)}}\,\, . \eqno\eqnlabel{nn8-b}$$
\suboff

{\settabs 8\columns
\+ \cr
\+ $D=3$ & $(K=16)$:\cr
\+\cr}

The ${\bf M}= {\bf M}^{4,4}$ model twisted by the simple current
$${\cal J}_{16}\equiv \phi_{K/4}^0\bar\phi_0^0\eqno\eqnlabel{j16}$$
producing,
\subon
$$\eqalignno{ V\left(K=16\right) &= \hbox to .38cm{\hfill}\vert (\t_{0,16} +
\t_{16,16}) +
(\t_{8,16} +\t_{-8,16})\vert^2_{untwisted}\cr
&\cr
&\hbox to 1em{\hfill}
+ \vert (\t_{4,16} + \t_{-4,16}) + (\t_{12,16} +\t_{-12,16})
\vert^2_{untwisted}\,\, .
 &\eqnlabel{nn16-a}}$$
Thus,
$$Z({\rm theta~factor,~} K=16)=\vert c^0_0 + c^0_8\vert^2_{_{(A_{16})}} +
4\vert c^0_4\vert^2_{_{(C_{16})}}\,\, .
\eqno\eqnlabel{nn16-b}$$
\suboff
(In this case the twisting is trivial since $J_{16}$ is in the initial
untwisted model.)

These simple twist currents are of the general form
$${\cal J}_K =
(\phi^0_{K/4})^{D-2}(\bar\phi^0_0)^{D-2}\,\, . \eqno\eqnlabel{gensc}$$
I believe this specific class of twist currents
is the key to space-time
supersymmetry in the parafermion models.\footnote[@]{$\bar {\cal J}_K=
(\phi^0_0)^{D-2}(\bar\phi^0_{K/4})^{D-2}$ is automatically generated as a
twisted state.} Without its twisting effect,
numerically zero fractional superstring MIPF's in three, four, and six
dimensions  cannot be formed.
This twisting also reveals much about the necessity of
non-$A_K$ sectors.  Terms from the twisted and untwisted
sectors of these models become equally mixed in the $\vert A_K\vert^2$,
$\vert B_K\vert^2$, and $\vert C_K\vert^2$ contribution to the level $K$
partition function.
Further, this twisting keeps the string functions with $n\not\equiv 0,K/2
\pmod{K}$ from mixing with those with $n\equiv 0,K/2 \pmod{K}$.  This is
especially significant since I believe the former string functions
in the $C_K$ sector
likely
correspond to space-time fields of fractional spin-statistics ({\it i.e.,}
anyons)
and the latter in both $A_K$ and $B_K$ to space-time bosons and
fermions.

Since in the antiholomorphic sector ${\cal J}_K$ acts as the identity, I will
focus on its effect in the holomorphic sector.
In the $A_K$ sector the operator
$(\phi_{K/4}^0)^{D-2}$ transforms the bosonic (fermionic)
nonprojection\footnote[$\$ $]{We use the same language as the authors of refs.
\putref{ref63,ref62,ref65}, \putref{ref12}.  Nonprojection refers to the
bosonic and fermionic fields in the $A^{boson}_K$ and $A^{fermion}_K$
subsectors, respectively, corresponding to string functions with positive
coefficients, whereas projection fields refer to those
corresponding to string functions with negative signs.  With this definition
comes an overall minus coefficient on $A_K^{fermion}$, as shown in eqn.
(2.31a).  For example,
in (2.31b), the bosonic non-projection fields are
\hbox {$(\phi^0_0 + \phi^2_0)^3(\phi^1_0)$}
 and the bosonic projection is
$(\phi^1_0)^4$. Similarly, in (2.31c) the fermionic non-projection
field is $(\phi^1_1)^4$ and the projections are $(\phi^0_1
+\phi^4_1)^3(\phi^1_1)$.}
fields into the
fermionic (bosonic) projection fields and vice-versa. For example, consider
the effect of this twist current on the fields represented in
\subon
$$ A_4\equiv A^{boson}_4 - A^{fermion}_4\,\, ,\eqno\eqnlabel{abosferm-a}$$
where
$$\eqalignno{A^{boson}_4 &= 4\left\{ (c^0_0 + c^4_0)^3(c^2_0) -
                                          (c^2_0)^4\right\}
                         &\eqnlabel{abosferm-b}\cr
             A^{fermion}_4 &= 4\left\{
                             (c^2_2)^4 - (c^0_2 + c^4_2)^3(c^2_2)\right\}\,\, .
                         &\eqnlabel{abosferm-c}\cr}$$
\suboff
Twisting by $(\phi^0_{K/4})^{D-2}$ transforms the related fields as
\subon
$$\eqalignno{ (\phi^0_0 + \phi^2_0)^3(\phi^1_0) &\hbox to 1cm{\hfill}
{\buildrel {(\phi^0_{K/4})^{D-2}}\over\Longleftrightarrow}\hbox to 1cm{\hfill}
(\phi^2_1 + \phi^0_1)^3 (\phi^1_1)
&\eqnlabel{phitwist-a}\cr (\phi^1_0)^4 &\hbox to 1cm{\hfill}
{\buildrel {(\phi^0_{K/4})^{D-2}}\over\Longleftrightarrow}\hbox to 1cm{\hfill}
(\phi^1_1)^4\,\, .
&\eqnlabel{phitwist-b}\cr}$$
\suboff

Although the full meaning of the projection fields is not yet understood,
the authors of refs. \putref{ref62} and \putref{ref12} argue that the
corresponding string functions
should be interpreted as ``internal'' projections, i.e.,
cancellations of degrees of freedom in the fractional superstring models.
Relatedly, the authors show that when the $A_K$ sector is
written as $A^{boson}_K - A^{fermion}_K$, as done above, the $q$-expansions of
both $A^{boson}_K$ and $A^{fermion}_K$ are all positive.  Including the
fermionic projection terms results in the identity
\subon
$$\eta^{D-2} A_K^{fermion} = (D-2)\left( {(\theta_2)^4\over
16\eta^4}\right)^{(D-2)/8}\,\, .
\eqno\eqnlabel{af-a}$$
Eqn. (\puteqn{af-a}) is the standard theta function expression for $D-2$
world sheet Ramond Majorana-Weyl fermions.
Further, $$\eta^{D-2} A_K^{boson} = (D-2)\left(
{(\theta_3)^4 - (\theta_4)^4\over 16\eta^4}\right)^{(D-2)/8}\,\, .
\eqno\eqnlabel{af-b}$$
\suboff

Now consider the $B_K$ sectors. For $K=4$ and $8$ the operator
$(\phi^0_{K/4})^{D-2}$ transforms the primary fields corresponding to the
partition functions terms in the first set of brackets on the RHS of eqns.
(\puteqn{part4-b},\puteqn{part8-b})
into the fields represented by the partition functions terms in the
second set. For example, in the $K=4$ ($D=6$) case
\subon
$$\eqalignno{(\phi^0_0 + \phi^2_0)(\phi^1_0)(\phi^0_1 + \phi^2_1)^2
&\hbox to 1cm{\hfill}
{\buildrel {(\phi^0_{K/4})^{D-2}}\over\Longleftrightarrow}\hbox to 1cm{\hfill}
(\phi^2_1 +
\phi^0_1)(\phi^1_1)(\phi^2_0+\phi^0_0)^2 & \eqnlabel{phib-a}\cr
(\phi^1_0)^2(\phi^1_1)^2 &\hbox to 1cm{\hfill}
{\buildrel {(\phi^0_{K/4})^{D-2}}\over\Longleftrightarrow}\hbox to 1cm{\hfill}
(\phi^1_1)^2(\phi^1_0)^2\,\, .&
\eqnlabel{phib-b}\cr}$$
\suboff

Making an analogy with what occurs in the $A_K$ sector, I suggest that
$(\phi^0_{K/4})^{D-2}$ transforms bosonic (fermionic)
nonprojection fields into fermionic (bosonic) projection fields and
vice-versa in the $B_K$ sector also.
Thus, use of the twist current ${\cal J}_K$ allows for bosonic and fermionic
interpretation of these
fields\footnote[$\& $]{Similar conclusions have been reached by K. Dienes and
P.
Argyres for different reasons. They have, in fact, found theta function
expressions for the $B_K^{boson}$ and $B_K^{fermion}$
subsectors.\markup{\putref{ref68}}}:
\subequationnumstyle{alphabetic}
$$B_4\equiv B^{boson}_4 - B^{fermion}_4\,\, ,\eqno\eqnlabel{bbf-a}$$ where
$$\eqalignno{B^{boson}_4 &= 4\left\{(c^0_0 + c^4_0)(c^2_0)(c^0_2+c^4_2)^2 -
(c^2_0)^2(c^2_2)^2\right\}&\eqnlabel{bbf-b}\cr B^{fermion}_4 &=
4\left\{(c^2_2)^2(c^2_0)^2 - (c^0_2+c^4_2)(c^2_2)(c^0_0
+c^4_0)^2\right\}\,\, .&\eqnlabel{bbf-c}\cr}$$
\subequationnumstyle{blank}
What appears as the projection term, $(c^2_0)^2(c^2_2)^2$, for the
proposed bosonic part acts as the nonprojection term for the fermionic half
when the subscripts are reversed.  One interpretation is
this implies a
compactification of two transverse dimensions.\footnote[$\#\# $]{This was also
suggested in ref. \putref{ref62} working from a different approach.} The
spin-statistics of the physical states of the $D=6$
model as observed in four-dimensional uncompactified space-time would then
be determined
by the (matching) $n$ subscripts of the first two string
functions\footnote[@@]{Using the subscripts $n'$ of last two string functions
to define spin-statistics in $D=4$ uncompactified space-time
corresponds to interchanging the definitions of $B^{boson}_4$ and
$B^{fermion}_4$.}
(corresponding to the two uncompactified transverse dimensions) in each term
of the form $c^{l_1}_n c^{l_2}_n c^{l_3}_{n'} c^{l_4}_{n'}$.  The $B_8$ terms
can
be interpreted similarly when one dimension is compactified.

However, the
$C_K$ sectors are harder to interpret. Under
$(\phi^0_{K/4})^{D-2}$ twisting, string functions with $K/4$ subscripts
are invariant, transforming back into themselves.  Thus, following
the pattern of $A_K$ and $B_K$ we would end up writing, for example,
$C_{16}$ as
\subequationnumstyle{alphabetic}
$$C_{16}= C^a - C^b \eqno\eqnlabel{cspin-a}$$ where, $$\eqalignno{C^a_{16}
&= (c^2_4 + c^{14}_4) - c^8_4 & \eqnlabel{cspin-b}\cr C^b_{16} &= c^8_4 -
(c^2_4 + c^{14}_4)\,\, . & \eqnlabel{cspin-c}\cr}$$
\subequationnumstyle{blank}

The transformations of the corresponding primary fields are not quite as
trivial, though.
$(\phi^1_2 + \phi^7_2)$ is transformed into its conjugate field
$(\phi^7_{-2} + \phi^1_{-2})$ and likewise $\phi^4_2$ into
$\phi^4_{-2}$.
suggesting that $C^a_{16}$ and $C^b_{16}$ are the partition functions
for conjugate fields. Remember, however, that $C_{16}=0$. Even though we may
interpret this sector as containing two conjugate space-time fields, this
(trivially) means that the partition function for each is identically zero.
I refer to
this effect in the $C_K$ sector as ``self-cancellation''. One
interpretation is that there are no states in the $C_K$ sector of the Hilbert
space that survive all of the internal projections. If this is correct,
a question may arise as to the consistency of the $K=8$ and
$16$ theories.  Alternatively, perhaps anyon statistics allow two
(interacting?)  fields of either identical fractional space-time spins
$s_1=s_2={2m\over K}$, or space-time spins related by $s_1={2m\over K}= 1-
s_2$,
where in both cases $0<m<{K\over 2} \pmod{1}$,
to somehow cancel each other's contribution to the partition function.

\sectionnumstyle{blank}
\section{\bf Conclusions}

A viable and consistent generalization of the superstring would be
significant. I have shown that the
partition functions for fractional superstrings have simple
origins when derived systematically
through the factorization approach of Gepner and Qiu.
Further, using this affine/theta function factorization of the
parafermion partition  functions, I have related the $A_K$ sector
containing the graviton and
gravitino with the massive sectors $B_K$ and $C_K$. A bosonic/fermionic
interpretation
of the $B_K$ subsectors was given.  Apparent ``self-cancellation'' of the
$C_K$
sector was shown. I am currently investigating the meaning of this last
issue.

\sectionnumstyle{blank}\subsectionnumstyle{blank}
\section{\bf Appendix: Simple and Non-Simple Currents}
\sectionnumstyle{Alphabetic}\sectionnum=1
\equationnum=0

A simple current $J_s$ is a primary field of a conformal field theory (CFT)
which, when fused with any other primary field (including itself)
$\Phi_{\bf l}$,
where $\bf l$ is a vector of indices labeling a primary field
of the CFT, produces only a single primary field as a product state:
$$J_s\otimes\Phi_{\bf l} = \Phi_{\bf l'}\, .\eqno\eqnlabel{app1}$$
A non-simple current
$J_{ns}$, when fused with at least one other primary field (possibly
itself), produces more than one term:
$$ J_{ns}\otimes\Phi_{\bf l}= \sum_{\bf l'} \Phi_{\bf l'}\,
.\eqno\eqnlabel{app2}$$

Schellekens {\it et al} \markup{\putref{ref17}} have shown that twisting
a known modular invariant model by a simple current results in a new
modular invariant.  Let $N$ be the order of the simple current. That is, $N$
is the smallest positive integer such that
$$ (J_s)^N\otimes\Phi_{\bf l} = \Phi_{\bf l}
\eqno\eqnlabel{app3}$$
for all primary fields $\Phi_{\bf l}$ in the CFT.  The
``untwisted sector'' of the new model for this CFT is formed by those primary
fields in the original, which when fused with any power of the current
$(J_s)^i$, $i=1 {\rm~to~} N-1$, only produce $T$ invariant states. The
``twisted sectors'' of the new model are formed by the sets of ($T$
invariant) fusion
products formed from each $(J_s)^i$ acting on the untwisted sector of the
new model.

Deriving new models using non-simple currents $J_{ns}$ is more difficult.
The general requirement that a primary field of the original model remain
in the untwisted sector of the twisted model appears to be that there be at
least one $T$ invariant in the fusion product of that field and $J_{ns}$.  The
simplest type of model involving non-simple currents is a class formed using
a closed subalgebra of non-simple currents $\{J_{ns,a}\}$.  For an
investigation into
models formed from non-simple twistings see ref. \putref{ref10}.

\sectionnumstyle{blank}
{\bf\section{References:}}

\begin{putreferences}
\reference{ref1}{M.~Green, J.~Schwarz, and E.~Witten,
\underline{Superstring Theory,} V. I \& II, (1987).\\
A.N.~Schellekens, ed. \underline{Superstring Construction,} (1989);\\
M.~Kaku, \underline{Strings, Conformal Fields and Topology}, (1991);\\
C.~Hull, {\it A Review of W Strings}, CTP TAMU-30/92 (1992).}
\reference{ref3}{F.~Ardalan and F.~Mansouri, {\it Phys. Rev.} {\bf D9} (1974)
3341;\\
\undbib {\it Phys. Rev. Lett.} {\bf 56} (1986) 2456;\\
\undbib {\it Phys. Lett.} {\bf B176} (1986) 99.}
\reference{ref4}{A. Zamolodchikov and V. Fateev, {\it Sov. Phys.} JETP
{\bf 62} (1985) 215;
\undbib {\it Teor. Mat. Phys.} {\bf 71} (1987) 163.}
\reference{ref5}{P.~Frampton and M.~Ubriaco, {\it Phys. Rev.} {\bf D38} (1988)
1341.}
\reference{ref61}{P.~Argyres, A.~LeClair, and S.-H.~Tye, {\it Phys. Lett.}
{\bf B235} (1991).}
\reference{ref62}{P.~Argyres and S.~-H.~Tye, {\it Phys. Rev. Lett.} {\bf 67}
(1991) 3339.}
\reference{ref63}{K.~Dienes and S.~-H.~Tye, {\it Nucl. Phys.} {\bf B376} (1992)
297.}
\reference{ref64}{P.~Argyres, J.~Grochocinski, and S.-H.~Tye, CLNS 91/1126.}
\reference{ref65}{P.~Argyres, K.~Dienes and S.-H.~Tye, CLNS 91/1113;
McGill-91-37.}
\reference{ref7}{D.~Gepner and Z.~Qiu, {\it Nucl. Phys.} {\bf B285} (1987)
423.}
\reference{ref8}{V.~Ka\v c, {\it Adv. Math.} {\bf 35} (1980) 264.\\
V.~Ka\v c and D.~Peterson, {\it Bull. AMS} {\bf 3} (1980) 1057;\\
\undbib {\it Adv. Math.} {\bf 53} (1984) 125.}
\reference{ref9} {H.~Kawai, D.~Lewellen, and S.-H.~Tye, {\it Nucl. Phys.}
{\bf B288} (1987) 1.\\ I.~Antoniadis, C.~Bachas, and C.~Kounnas, {\it Nucl.
Phys.} {\bf B289} (1987) 87.}
\reference{ref10}{G.~Cleaver and D.~Lewellen, CALT-68-1754.}
\reference{ref11} {Wilczek, {\it Fractional Statistics and Anyon
Superconductivity.}}
\reference{ref12} {P.~Argyres, E.~Lyman, and S.-H.~Tye, CLNS 91/1121.}
\reference{ref13} {H.S.~Green, {\it Phys. Rev.} {\bf 90} (1953) 270.}
\reference{ref14} {F.~Mansouri and X.~Wu, {\it Mod. Phys. Lett.} {\bf A2}
(1987) 215;\\
\undbib {\it Phys. Lett.} {\bf B203} (1988) 417;\\
\undbib {\it J. Math. Phys.} {\bf 30} (1989) 892;\\
A.~Bhattacharyya {\it et. al.}, {\it Mod. Phys. Lett.} {\bf A4} (1989) 1121.\\
\undbib {\it Phys. Lett.} {\bf B224} (1989) 384.}
\reference{ref15} {I.~Antoniadis and C.~Bachas, {\it Nucl. Phys.} {\bf
B278} (1986) 343;\\
M.~Hama, M.~Sawamura, and H.~Suzuki, RUP-92-1.}
\reference{ref16} {K.~Li and N.~Warner, {\it Phys. Lett.} {\bf B211} (1988)
101;\\
A.~Bilal. {\it Phys. Lett.} {\bf B226} (1989) 272;\\
G.~Delius. ITP-SB-89-12.}
\reference{ref17} {A.N.~Schellekens and S.~Yankielowicz, {\it Nucl. Phys.}
{\bf B327} (1989) 3;\\
A.N.~Schellekens, {\it Phys. Lett.} {\bf 244} (1990) 255;\\
B.~Gato-Rivera and A.N.~Schellekens, {\it Nucl. Phys.} {\bf B353} (1991) 519.\\
\undbib {\it CERN-TH.6056/91.}}
\reference{aspects}{G.~Cleaver and P.~Rosenthal, CALT-68-1756.}
\reference{ref68}{K.~Dienes, Private communications.}
\reference{me}{G.~Cleaver, Unpublished research.}
\reference{thesis}{G.~Cleaver, Ph.D. Thesis.}
\end{putreferences}

\bye